\documentclass[usenatbib]{mn2e}
\usepackage{apjfonts}
\usepackage{epsfig}
\usepackage{amssymb}
\usepackage{amsmath}
\usepackage{fixltx2e} 
\usepackage{deluxetable}

\newcommand{\scaleup}{}

\newcommand{\plotside}[1]
 {\centering \leavevmode \includegraphics[width={0.95\textwidth}]{#1}}
\newcommand{\acknowledgments}{\begin{small}\section*{Acknowledgments}\end{small}}
\newcommand\altaffilmark[1]{$^{#1}$}
\newcommand\altaffiltext[1]{$^{#1}$}
\newcommand{\tableclear}{}

\newcommand{\etal}{et al.}

\newcommand{\msun}{M_{\sun}}

\title[A Maximum Stellar Surface Density]{A Maximum Stellar Surface Density in
  Dense Stellar Systems}

\author[Hopkins \etal]{
\parbox[t]{\textwidth}{ 
Philip F.~Hopkins,\thanks{E-mail:phopkins@astro.berkeley.edu}\altaffilmark{1} 
Norman Murray,\altaffilmark{2,3} 
Eliot Quataert,\altaffilmark{1} \& 
Todd A.~Thompson\altaffilmark{4,5,6}}\vspace*{6pt} \\
\altaffiltext{1}{Department of Astronomy and Theoretical Astrophysics Center, University of California Berkeley, Berkeley, CA 94720} \\
\altaffiltext{2}{Canadian Institute for Theoretical Astrophysics, 
60 St.\ George Street, University of Toronto, ON M5S 3H8, Canada} \\
\altaffiltext{3}{Canada Research Chair in Astrophysics} \\
\altaffiltext{4}{Department of Astronomy, The Ohio State University, 
140 W.\ 18th Ave., Columbus, OH 43210} \\
\altaffiltext{5}{Center for Cosmology \&\ Astro-Particle Physics, 
The Ohio State University, 191 W.\ Woodruff Ave., 
Columbus, OH 43210} \\
\altaffiltext{6}{Alfred P.~Sloan Fellow} }

\date{Submitted to MNRAS, August 10, 2009}
\begin{document}
\maketitle
\label{firstpage}

\begin{abstract}

  We compile observations of the surface mass density profiles of
  dense stellar systems, including globular clusters in the Milky Way
  and nearby galaxies, massive star clusters in nearby starbursts,
  nuclear star clusters in dwarf spheroidals and late-type disks,
  ultra-compact dwarfs, and galaxy spheroids spanning the range from
  low-mass, ``cusp'' bulges and ellipticals to massive ``core''
  ellipticals.  We show that in all cases the maximum stellar surface
  density attained in the central regions of these systems is similar,
  $\Sigma_{\rm max}\sim 10^{11}\,\msun\,{\rm kpc^{-2}}$ ($\sim 20$ g
  cm$^{-2}$), despite the fact that the systems span $\sim 7$ orders
  of magnitude in total stellar mass $M_\ast$, $\sim 5$ in effective
  radius $R_{e}$, and have a wide range in {\em effective} surface
  density $M_{\ast}/R_{e}^{2}$.  The surface density limit is reached
  on a wide variety of physical scales in different systems and is
  thus not a limit on three-dimensional stellar density.  Given the
  very different formation mechanisms involved in these different
  classes of objects, we argue that a single piece of physics likely
  determines $\Sigma_{\rm max}$.  The radiation fields and winds
  produced by massive stars can have a significant influence on the
  formation of both star clusters and galaxies, while neither
  supernovae nor black hole accretion are important in star cluster
  formation.  We thus conclude that feedback from massive stars likely
  accounts for the observed $\Sigma_{\rm max}$, plausibly because star
  formation reaches an Eddington-like flux that regulates the growth
  of these diverse systems.  This suggests that current models of
  galaxy formation, which focus on feedback from supernovae and active
  galactic nuclei, are missing a crucial ingredient.

\end{abstract}


\section{Introduction}
\label{sec:intro}

Feedback from massive stars and black holes is widely believed to play
a critical role in the formation of other stars, stellar clusters, and
entire galaxies.  However, the precise physical mechanism(s) that
dominate the coupling between stars and/or black holes and their
surrounding environment have not been definitively established.
Unlike the energy deposited by, e.g., a jet, the momentum produced by
a central star or accreting black hole cannot be radiated away.  When
the momentum deposition rate exceeds the strength of gravity, gas can
be efficiently expelled from the system.  This limit is analogous to
the Eddington limit from stellar physics; however, the gas in
star-forming regions and galaxies is dusty, and thus the appropriate
opacity is not that of electron scattering, but the much larger
opacity due to dust \citep[see
e.g.][]{scoville:2001.dust.pressure.in.sb.regions,
  murray:momentum.winds, thompson:rad.pressure}.

This generalization of the Eddington limit has been posited as an
explanation for the characteristic sizes of massive stellar clusters
\citep{murray:sizes.lum.star.clusters}, as well as the maximum
luminosities of the most rapidly star-forming objects in the Universe,
bright sub-millimeter galaxies with $\sim10^{13}\,L_{\sun}$ inside of
$\sim$kpc \citep[star formation rates $\gtrsim1000\,\msun\,{\rm
  yr^{-1}}\,{\rm kpc^{-2}}$; e.g.,][]{younger:smg.sizes,
  walter:2009.hyperlirg.in.highz.qso.host}.  Outflows
driven by radiation pressure may help explain phenomena as diverse as
turbulence in star-forming regions, the relation between galaxy mass
and stellar population metallicity, the correlations between the
masses of black holes and their host spheroids, and the abundance
patterns of the IGM \citep[see e.g.][]{murray:momentum.winds,
  thompson:rad.pressure, oppenheimer:outflow.enrichment,
  oppenheimer:metal.enrichment.momentum.winds,
  krumholz:rad.pressure.fx.on.sf,murray:molcloud.disrupt.by.rad.pressure,
  krumholz:2009.rad.dom.region.dynamics}.  
Small changes in the nature of
stellar feedback --- for example, whether or not winds are really
momentum or pressure driven, or the characteristic scales of wind
driving --- have dramatic consequences for predictions of models of galaxy
formation, affecting the global structure of even Milky Way-mass
systems \citep{governato:disk.formation,scannapieco:fb.disk.sims}.

Outflows appear to be ubiquitous from rapidly star-forming galaxies
\citep{heckman:1990.sb.superwinds,martin99:outflow.vs.m,martin05:outflows.in.ulirgs,
  heckman:superwind.abs.kinematics,
  erb:lbg.metallicity-winds,weiner:z1.outflows}.  However, quantifying
the basic properties of these outflows is challenging; their
mass-loading, dynamical importance, and implications for future star
formation remain subjects of considerable debate. This uncertainty
translates into uncertainty in the dominant physical processes driving
such outflows (e.g., radiation pressure, supernovae, AGN vs. star
formation, etc.).  It is also unclear whether or not feedback
processes and the outflows they generate actually leave an imprint on
the structural properties of star clusters and galaxies.

In this {\it Letter}, we show that a diverse set of dense stellar
systems --- from star clusters to the centers of elliptical galaxies
--- have a maximum stellar surface density. We argue that this
observational fact provides a strong constraint on models of feedback
in star cluster and galaxy formation, pointing to the critical role of
massive stars, rather than supernovae or AGN.
In \S~\ref{sec:obs}, we present our compilation of observations of
dense stellar systems (Table~\ref{tbl:obs}), and compare their surface
density profiles and maximum stellar surface densities.  
In \S~\ref{sec:discussion}, we discuss the implications of our results
and interpret the observed maximum stellar surface density in terms of
models in which star formation reaches the (dust) Eddington limit.

Throughout, we adopt a $\Omega_{\rm M}=0.3$, $\Omega_{\Lambda}=0.7$,
$h=0.7$ cosmology and a \citet{chabrier:imf} stellar IMF, but these
choices do not affect our conclusions. Changes in the IMF would
systematically shift the stellar masses of the massive systems, but
reasonable (factor $<2$) changes are within the scatter in the data we
compile.

\vspace{-0.3in}
\section{Observational Results}
\label{sec:obs}

In Table~\ref{tbl:obs}, we summarize our compilation of a large sample
of the observed stellar mass surface density profiles of dense stellar systems,
including globular clusters (GCs) in the Milky Way and nearby
galaxies,\footnote{We do not distinguish between GCs that might have
  undergone core collapse and those that have not (see \S~\ref{sec:discussion}).} the nuclear
stellar disk observed around Sgr A$^{\ast}$ in the galactic center,
massive star clusters in starburst regions, nuclear star clusters in
dwarf spheroidals and late-type disks, ultra-compact dwarfs, and
classical spheroids spanning the range from low-mass, ``cusp'' bulges
and ellipticals (those with steep central surface brightness slopes)
to massive ``core'' galaxies (these predominate at the highest masses,
with shallower central slopes).  We also include massive compact
ellipticals observed at high redshifts, $z\sim2-3$; as already shown
in \citet{hopkins:density.galcores} and
\citet{bezanson:massive.gal.cores.evol}, these compact high $z$
systems in fact have the same maximum density as the central regions
of today's ellipticals.

{\textwidth 3.0in 
\tableclear
\begin{deluxetable}{lll}
\tablecolumns{3}
\tabletypesize{\scriptsize}
\tablecaption{Observations of Dense Stellar Systems\label{tbl:obs}}
\tablewidth{0pt}
\tablehead{
\colhead{Reference\tablenotemark{1}} &
\colhead{Description\tablenotemark{2}} &
\colhead{Symbol\tablenotemark{3}} 
}
\startdata
\citet{lu:mw.nuclear.disk} & MW Nuclear Disk & dark green square
\\
\citet{harris96:mw.gcs} & MW GCs & pink circle
\\
\citet{barmby:m31.gcs} & M31 GCs & yellow circle
\\
\citet{rejkuba:cenA.gcs} & Massive Cen A GCs & green pentagon
\\
\citet{mccrady:m82.sscs} & M82 SSCs & orange $\times$
\\
\citet{walcher05:nuclei.mdyn} & Sd+dSph nuclear SCs & red $+$
\\
\citet{boker04:nuclei.scalings} & Sd disk nuclear SCs & magenta asterisk
\\
\citet{geha02:dE.nuclei} & dSph nuclei & violent inv.\ triangle
\\
\citet{hasegan:M87.ucds} & M87 UCDs & light blue star
\\
\citet{evstigneeva:virgo.ucds} & Virgo UCDs & dark blue triangle
\\ 
\citet{hilker:fornax.ucds} & Fornax UCDs & cyan diamond
\\
\citet{jk:profiles} & Virgo Es & red circle
\\
\citet{lauer:bimodal.profiles} & local massive Es & violet square
\\
\citet{vandokkum:z2.sizes} & $z=2-3$ massive Es & orange star\\
\enddata
\tablenotetext{1}{For most of the objects we adopt the fitted
  functional forms for the density profiles presented in each
  paper. For the \citet{jk:profiles} and
  \citet{lauer:bimodal.profiles} samples, however, the data are
  sufficiently high dynamic range that we use the observed profiles
  directly. Note that the profiles shown in
  \citet{lauer:bimodal.profiles} include only the HST data; the
  profiles used here are the composite (HST+ground-based) profiles
  used therein to determine effective radii.}
\tablenotetext{2}{GC$=$Globular cluster, SC$=$Star cluster,
  SSC$=$Super star cluster, UCD$=$Ultra-compact dwarf,
  E$=$elliptical. Note that \citet{geha02:dE.nuclei} and
  \citet{walcher05:nuclei.mdyn} refer to their dwarf galaxy hosts of
  nuclear star clusters as dEs; we  follow \citet{jk:profiles}
  and classify these as dwarf spheroidals (dSph).}  \tablenotetext{3}{Plotting symbol used for
  objects in each sample throughout.}
\end{deluxetable}
\tableclear
}

For each object, we adopt the best-fit surface brightness profile
determined by the authors. This is usually a \citet{king:profile}
profile for the less massive systems, and a \citet{sersic:profile} or
Nuker profile for the more massive objects.  For the MW nuclear
stellar disk we adopt the total mass from \citet{lu:mw.nuclear.disk}
and the profile shape fitted in \citet{paumard:mw.nuclear.disk}
($\Sigma_{\ast}\propto (R^{2}+R_{\rm core}^{2})^{-1}$).\footnote{We use the 
notation $\Sigma_{\ast}$, $\Sigma_{\rm gas}$, and $\Sigma_{\rm tot}$ 
to refer to the stellar, gas, and total (stellar plus gas plus dark matter) 
surface densities. Throughout, the terms gas mass or gas surface density refer specifically to
``cold'' (i.e.\ star-forming or rotationally supported disk) gas. } 
The data from
\citet{jk:profiles} and \citet{lauer:bimodal.profiles} span sufficient
dynamic range that we directly adopt the observed (PSF de-convolved)
profiles rather than a fit to the data; for the other systems in Table
{\ref{tbl:obs}, the data do not span sufficient dynamic range to see
  significant deviations from the fits used here.  In all cases, we
  convert the measured light profiles to stellar mass density profiles
  by first adopting the authors' determination of total stellar masses
  (re-normalized to our \citealt{chabrier:imf} IMF where
  appropriate).  We then assume a radius-independent stellar
  mass-to-light ratio to convert the light profiles to mass profiles.
  The assumption of a radius-independent $M_{\ast}/L$ is reasonable
  for the well-resolved galactic systems (nuclear SCs and Es), which
  are relatively old and have weak observed color or stellar
  population gradients \citep[see e.g.][]{cote:virgo,
    sanchezblazquez:ssp.gradients,hopkins:cusps.fp}. The lower-mass
  systems (e.g.\ GCs) are approximately single stellar populations, so
  this should be a reasonable first approximation in those cases as
  well.  In the high $z$ ellipticals the central densities are
  unresolved; we thus extrapolate the best-fit Sersic profiles
  inwards. For low-redshift ellipticals, this extrapolation is
  typically a factor $\sim2$ higher than the true central densities
  (see \citealt{hopkins:density.galcores} for more details).

\begin{figure*}
    \centering
    \scaleup
    %\plotside{max_densities_fp_rev.ps}
    \plotside{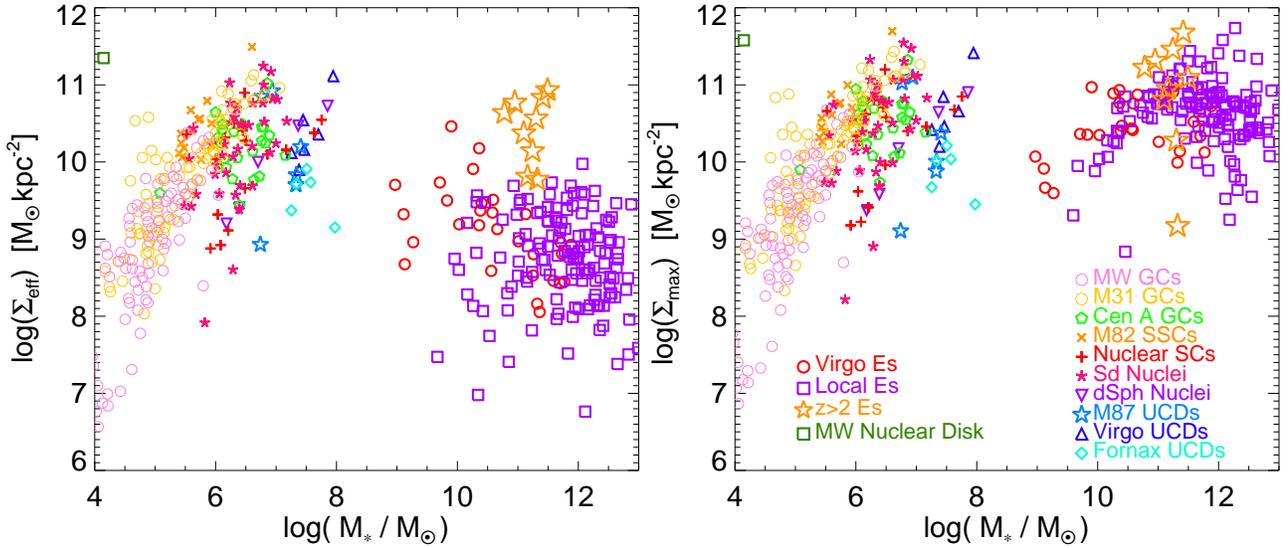}
    \caption{{\em Left:} Effective surface stellar mass density $\Sigma_{\rm
        eff} \equiv M_\ast(<R_{e})/\pi R_e^2$ as a function of total stellar
      mass for the different samples described in the text and legend.
{\em Right:}  Maximum  
surface stellar mass density $\Sigma_{\rm max}$ obtained at small radii in the 
surface density profile (see Fig. \ref{fig:profiles}), as a function of total stellar mass, for the different samples described in the text and legend.   The {effective} densities vary significantly across the different samples.  However, there appears to be a well-defined upper limit
$\sim 10^{11}\,\msun\,{\rm kpc^{-2}}$ to the maximum 
surface stellar mass density; this limit holds for the full range of dense stellar systems considered here.
\label{fig:mass.radius}}
\end{figure*}

The left panel of Figure~\ref{fig:mass.radius} shows the {\it
  effective} surface stellar mass densities $M_\ast/\pi R_e^2$ of these dense
stellar systems as a function of their total stellar mass, where
$R_{e}$ is the projected half-stellar mass radius.
The wide range of systems shown in Figure \ref{fig:mass.radius} differ
significantly in their large-scale structure and formation mechanisms.
Various works have compared these different object classes in this and
other projections of the fundamental plane, e.g., in the space of
effective surface or luminosity density, mass, mass-to-light ratio, size, or
velocity dispersion \citep{kormendy:spheroidal1,
  geha02:dE.nuclei,hasegan:M87.ucds,
  rejkuba:cenA.gcs,evstigneeva:virgo.ucds,dabringhausen:clusters.vs.gal.on.fp,
  hopkins:cusps.ell}.  In some of these correlations
there is apparent continuity from star clusters to massive galaxies,
but in others, different objects trace nearly perpendicular
correlations and thus appear physically quite distinct.
In terms of their {effective} densities, Figure \ref{fig:mass.radius}
shows that massive galaxies are significantly less dense than low-mass
stellar clusters, and there is no indication of a universal maximum
stellar surface density.

However, $\Sigma_{\ast}$ is of course higher at small $R\ll R_{e}$.
Figure~\ref{fig:profiles} thus directly compares the surface stellar mass
density profiles of these different systems. For clarity, rather than
plotting every individual profile, we plot the median and
$\pm1\,\sigma$ scatter in $\Sigma_{\ast}(R)$ at each $R$ for each class of
objects listed in Table~\ref{tbl:obs} (we neglect the low-mass GCs,
which have much lower densities in Figure~\ref{fig:mass.radius} and
remain relatively low-density at all radii).  Since the elliptical
samples, in particular that of \citet{jk:profiles}, span a very large
range in $R_{e}$ and $M_{\ast}$, we split them into 3 classes:
low-mass ellipticals in Virgo ($M_{\ast}<10^{10}\,\msun$) and high
mass cusp and core ellipticals.

\begin{figure*}
    \centering
    \scaleup
    %\plotside{compare_profiles_vsclass.ps}
    \plotside{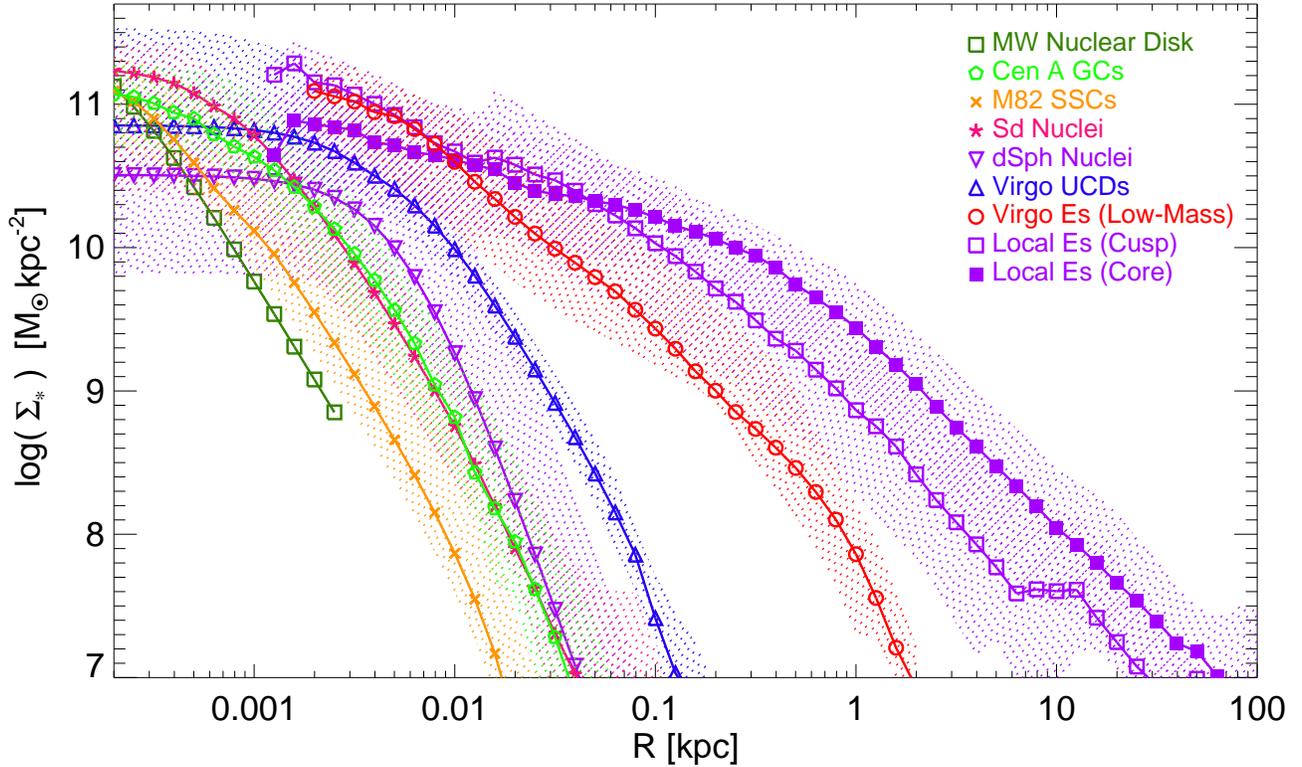}
    \caption{Stellar mass surface density profiles for the different samples. 
    For each object class, thick solid lines (with points) show the median 
    stellar mass surface density profile; shaded range shows the 
    $\pm1\,\sigma$ dispersion. 
    The upper limit $\sim 10^{11}\,\msun\,{\rm kpc^{-2}}$ is apparent, across a variety of classes. 
%\vspace*{2cm}
    \label{fig:profiles}}
\end{figure*}

Figure~\ref{fig:profiles} shows that the stellar mass surface density profiles
appear to asymptote to a maximum $\Sigma_{\rm max}
\sim10^{11}\,\msun\,{\rm kpc^{-2}} \sim 20$ g cm$^{-2}$, {\em
  independent of object class}, despite spanning an enormous range of
$M_\ast$ and $R_e$. The most massive Es maintain this density out to
few $100$\,pc scales, with $\Sigma_{\ast}$ decreasing only weakly with
radius.  Massive GCs, on the other hand, also reach this surface
density at small radii, but their densities then fall off rapidly with
radius.  Note that radii $\ll$pc are not resolved and are thus not
plotted for these systems; extrapolations of the King model fits to
arbitrarily small $R$, however, do not significantly exceed the
maximum densities shown in Figure~\ref{fig:profiles}.  Even the MW
nuclear disk, with its proximity to the nuclear black hole and
evidence for a top-heavy IMF
\citep{nayakshin:2005.top.heavy.imf.mw.center}, does not appear to
exceed $\Sigma_{\rm max}$.  Interestingly, recent observations suggest
that there is a nuclear star cluster in the center of the Milky Way.
The observations favor a radius-independent stellar surface density
$\approx1.5-3\times 10^{11}\,\msun\,{\rm kpc^{-2}}$ from $\sim
0.02-1.0$\,pc \citep{schoedel:mw.nuclear.cluster}; this is comparable
to $\Sigma_{\rm max}$ found here.

The right panel of Figure~\ref{fig:mass.radius} shows the {\em
  maximum} stellar surface density $\Sigma_{\rm max}$ of each object
as a function of its total stellar mass. For the King profile objects,
$\Sigma_{\rm max}$ is defined to be the stellar surface density inside the
fitted ``core'' radius ($\sim R_{e}$).  For the ellipticals, we define
$\Sigma_{\rm max}$ as the $\langle \Sigma_{\ast} \rangle$ inside the radius
where $\Sigma_{\ast}$ falls a factor $\sim2$ below the maximum $\Sigma_{\ast}$
measured.\footnote{For core ellipticals, the maximum 
is well-defined. We caution, however, that for cusp ellipticals, it is not 
yet established whether or not the profile continues to rise at sub-pc 
radii. At these radii, though, the profiles are not necessarily representative 
of those at formation, as $N$-body relaxation and core-collapse 
can occur. We therefore 
focus on the maxima at the range of fixed small radii observed ($\sim1-10\,$pc), 
which are consistent (see Figure~\ref{fig:profiles}). }
  Reasonable variations in our definition of $\Sigma_{\rm
  max}$ make little difference to our general conclusion: although the
effective densities of dense stellar systems (left panel) vary widely,
many such systems (and most of the object classes) have roughly the
same maximum stellar surface density $\sim 10^{11}\,\msun\,{\rm
  kpc^{-2}}$ at small radii, over a factor of $\sim 10^7$ in $M_\ast$.

\vspace{-0.15in}
\section{Discussion and Interpretation}
\label{sec:discussion}

The fact that a wide range of star clusters, dwarf galaxies, and
massive galaxies have the same maximum stellar surface density
$\Sigma_{\rm max} \sim10^{11}\,\msun\,{\rm kpc^{-2}} \sim 20$ g
cm$^{-2}$ suggests that a common physical process operated during the
formation of these diverse systems to limit the stellar density that
could be attained.  It is worth emphasizing that the observed maximum
is a maximum {\em surface} density. The different radii and masses
sampled here imply that the same stellar surface density corresponds to very
different {\em three-dimensional} stellar mass densities $\rho$.
Whatever process regulates the growth of these systems selects a
particular surface density, {\em not} a particular three-dimensional
density. 

It is, in principle, possible that the observed $\Sigma_{\rm max}$ is
due to the absence of sufficient gas reaching surface densities $\gg
\Sigma_{\rm max}$.  For example, in galaxy mergers, the high star
formation efficiency in pre-merger disks with high gas fractions and
masses, as well as the inefficient angular momentum loss during
gas-rich mergers, mean that it is difficult to
generate gas surface densities far in excess of $\Sigma_{\rm max}$ on
scales $\gtrsim 100$ pc
\citep[see e.g.][]{hopkins:disk.survival}; this same conclusion may
not, however, hold on smaller scales where $\Sigma_{\ast}$ is nonetheless
observed to be $\lesssim \Sigma_{\rm max}$
(Fig. \ref{fig:profiles}). In the case of star clusters, the masses of
the most massive giant molecular cloud (GMC) complexes present a
similar limitation; on $\sim1-10\,$pc scales, gas surface densities
$\gg\Sigma_{\rm max}$ would require GMC masses in excess of
$\sim10^{7}-10^{8}\,\msun$, at the limit of those inferred
observationally.

The observed maximum surface density occurs, however, over a huge
dynamic range in mass ($\sim 10^6-10^{12}\,\msun$), effective radius
($\sim3\,{\rm pc} - 50\,{\rm kpc}$), stellar population age (young
clusters with ages $\sim 10^{6}\,$yr to spheroids and globular
clusters with ages $\gtrsim10\,$Gyr) and other properties.  The
formation mechanisms and formation timescales of these different
classes of objects are also very different.  Whatever determines their
maximum stellar surface density cannot, therefore, be specific to details of
their formation or their global properties (e.g., $M_\ast$, $R_e$,
etc.). As a result, it requires significant fine-tuning for the same stellar 
$\Sigma_{\rm max}$ to apply independently, due to the availability of
gas, in such a wide range of systems.  Instead, it appears likely that
there is some rather generic physics that sets $\Sigma_{\rm max}$. The
key traits the systems we consider share are that they are
baryon-dominated and likely formed in dissipational 
(rapid, gas-dominated) events. 

%Because star clusters form and disrupt on
%a timescale much less than the time for massive stars to explode as supernovae
%(e.g., \citealt{murray:molcloud.disrupt.by.rad.pressure}), and because the time for stars to
%form at a gas surface density 
%$\Sigma_{\rm gas} \sim \Sigma_{\rm max}$ is also less than the time
%for massive stars to explode,\footnote{Note that the dynamical timescale is 
%$t_{\rm dyn}\sim(G\Sigma_{\rm max}/r)^{-1/2}\sim1.5\times10^5r_{10}^{1/2}$\,yr,
%where $r_{10}=r/10$\,pc.} it is unlikely that supernova feedback
%is important in setting $\Sigma_{\rm max}$.  
%Even if global star formation proceeds more slowly, the input energy from 
%supernovae scales with stellar mass (for a typical IMF with $\sim 10\%$ of the 
%stellar mass in massive stars, each of which explodes with $10^{51}\,{\rm erg}$, 
%$E_{\rm SN}\sim (10^{49}\,{\rm erg\,\msun^{-1}})\,\Sigma_{\rm \ast}\,R^{2}$). Compare 
%the binding energy, which scales $E_{\rm bind}\sim G\,M^{2}/R \sim G\,\Sigma_{\rm tot}^{2}\,R^{3}$. 
%As surface densities increase, binding energy increases $\propto \Sigma_{\rm tot}^{2}$, 
%faster than either the stellar mass surface density or star formation rate surface density. 
%The same occurs if one considers the rate of energy input ($\propto \dot{\Sigma}_{\ast}$) 
%versus $E_{\rm bind}/t_{\rm dyn}$. 
%Energetic feedback thus does not naturally set a $\Sigma_{\rm max}$. 

Because star clusters form and disrupt on a timescale much less than
the time for massive stars to explode as supernovae (e.g.,
\citealt{murray:molcloud.disrupt.by.rad.pressure}), and because the
time for stars to form at a gas surface density $\Sigma_{\rm gas} \sim
\Sigma_{\rm max}$ is also less than the time for massive stars to
explode,\footnote{Note that the dynamical timescale is $t_{\rm
   dyn}\sim(G\Sigma_{\rm
   max}/r)^{-1/2}\sim1.5\times10^5r_{10}^{1/2}$\,yr, where
 $r_{10}=r/10$\,pc.} it is unlikely that supernova feedback is
important in setting $\Sigma_{\rm max}$.  In addition, the energy
input from supernovae scales as $E_{\rm SN}\sim (10^{49}\,{\rm
 erg\,\msun^{-1}})\,\Sigma_{\rm \ast}\,R^{2}$, while the binding
energy of gas scales as $E_{\rm bind}\sim G\,M^{2}/R \sim
G\,\Sigma_{\rm gas} \Sigma_{\rm tot}\,R^{3}$.  Although energy
feedback can in principle pick out a specific $\Sigma_{\rm max}$, it
is only {\it below} such a surface density, not {\it above} it, that
feedback from energy deposition is important; this is contrary to the
phenomenology suggested by the data.

Similarly, there is no
evidence for energetically important black hole accretion during star
cluster formation.  The similarity of $\Sigma_{\rm max}$ in both star
clusters and galaxies thus suggests that massive stars themselves (on
or near the main sequence) are critical for setting the observed
$\Sigma_{\rm max}$ in dense stellar systems.  Although we do not have
a fully satisfactory theoretical explanation for $\Sigma_{\rm max}$ in
the context of this interpretation, we briefly describe the possibility
that it is set when these systems reach the dust Eddington limit.

The Eddington luminosity ($L_{\rm Edd}$) or flux ($F_{\rm Edd}$) for
dusty gas is
\begin{equation}
\frac{L_{\rm Edd}}{M_{\rm tot}}=\frac{F_{\rm Edd}}{\Sigma_{\rm tot}}=\frac{4\pi Gc}{\kappa_F}
\approx2500\,\,\kappa_{10}^{-1}\,\,{\rm ergs\, \, s^{-1} \,\, g^{-1}},
\label{edd1}
\end{equation}
where $\kappa_{10}$ is the opacity in units of 10\,cm$^2$ g$^{-1}$,
$M_{\rm tot}$ and $\Sigma_{\rm tot}$ are the total enclosed {\em total} (gas, stars, 
and dark matter) mass and
surface density, and $\kappa_F$ is the flux-mean dust opacity.  When
the medium is optically-thick to the re-radiated FIR emission
($\Sigma_{\rm gas}\gtrsim0.5$\,cm$^2$ g$^{-1}$) $\kappa_F$ can be approximated
by the Rosseland-mean dust opacity, which has the form
$\kappa_R\approx\kappa_0 T^2$ for $T\lesssim200$\,K and
$\kappa_R\sim5-10$\,cm$^2$ g$^{-1}\approx$\,\,constant for $200$\,K
$\lesssim T\lesssim T_{\rm sub}$, where
$\kappa_0\approx2\times10^{-4}$\,cm$^2$ g$^{-1}$ K$^{-2}$ and $T_{\rm
  sub} \sim 1500$ K is the sublimation temperature of dust
\citep{semenov:2003.dust.opacities}.  The surface densities in Figures
\ref{fig:mass.radius} \& \ref{fig:profiles} are sufficiently high that
if the systems had gas fractions of order $f_g\sim0.01$ or larger
during formation, then the medium was optically-thick in the FIR
($\kappa_Rf_g\Sigma_{\rm max}>1$).  For gas fractions of order unity,
the temperature deep inside the optically thick gas was $T\gtrsim200$
K and $\kappa_R\sim5-10$\,cm$^2$ g$^{-1}$ (for solar metallicity and
Galactic gas-to-dust ratio).  These considerations motivate our
scaling for $\kappa$ in the last equality of equation (\ref{edd1}).

We can compare the Eddington flux to the flux from star formation
implied by a given gas surface density if we extrapolate the observed
Kennicutt-Schmidt relation. From \citet{kennicutt98},
$\dot{\Sigma}_{\ast} \approx 2.5\times10^{-4}\,\msun\,{\rm
  yr^{-1}\,kpc^{-2}}\, [\Sigma_{\rm gas}/\msun\,{\rm pc^{-2}}]^{1.5}$;
other studies of high surface density systems suggest the index may be
somewhat steeper, $\sim1.7$ \citep{bouche:z2.kennicutt}.  Together
with the empirically calibrated relation between SFR and bolometric or
total infrared luminosity, $L_{\rm SF} = 2.22\times10^{36}\,{\rm W}\,
(\dot{M}_{\ast} / M_{\sun}\,{\rm yr^{-1}})$, this yields a simple
scaling of flux from young stars versus surface gas density;
\begin{equation}
\frac{F_{\rm SF}}{\Sigma_{\rm gas}} \approx 2.8\,{\rm ergs\,\,s^{-1}\,\,g^{-1}}\,
{\Bigl (}\frac{\Sigma_{\rm gas}}{10^{6}\,\msun\,{\rm kpc^{-2}}}{\Bigr)}^{0.5-0.7}\ 
\label{schmidt}
\end{equation}
This relation shows that the flux from star formation reaches and then
exceeds the threshold set by equation~(\ref{edd1}) for $\Sigma_{\rm
  gas}\sim10^{11}-10^{12}\,\msun\,{\rm kpc^{-2}}$ (depending on the
index adopted), suggestively similar to the maximum we find.  The
Eddington flux can indeed be produced by a young stellar population
($\lesssim10^{6.6}$\,yr), for which the stellar light to mass ratio is
$\Psi \equiv F_{\ast}/\Sigma_{\ast} \simeq 3000$ ergs s$^{-1}$ g$^{-1}$ for a
standard IMF.  A young stellar population is thus just capable of
reaching the dust Eddington limit and supporting an optically-thick
self-gravitating medium with radiation pressure.  In this limit,
however, it is unclear whether stellar radiation pressure can account
for a maximum surface stellar density $\Sigma_{\rm max}$ since both $F_{\ast}
\propto \Sigma_{\ast}$ and $F_{\rm edd} \propto \Sigma_{\rm tot}$.

Another possibility is that the relevant luminosity could, for a short
period of time, come directly from the inflow of gas required to form
the system, rather than from starlight.  If a gas-dominated system of
mass $M$ and gas surface density $\Sigma_{\rm gas}\sim \Sigma_{\rm tot}$ 
collapses on a free-fall time,
the luminosity due to the release of gravitational binding energy is
$L \simeq \dot M v_{ff}^2 \propto \Sigma_{\rm gas}^{5/4} M^{1/4}$, where $\dot M
\simeq M/[R/v_{ff}]$ is the inflow rate and $v_{ff}$ is the free-fall
velocity.  Comparing the inflow luminosity with the Eddington
luminosity, it is straightforward to show that there is a critical
gas surface density above which the inflow luminosity exceeds the
Eddington luminosity: $\Sigma_{\rm crit} \approx 3\times10^{11} \,
{M_{\odot}}{\rm kpc^{-2}} \, \kappa_{10}^{-4/5}\,M_8^{1/5}$, where
$M_8=M/10^8$\,M$_\odot$ is the mass at $\Sigma_{\rm gas} \sim \Sigma_{\rm
  crit}$, which is not necessarily equal to the total mass of the
system.  This estimate is intriguing both because it is close to our
observationally-inferred maximum surface density and because it
provides a clear mechanism for obtaining a {\it maximum} in the gas
(and ultimately stellar) surface density: the flux produced by inflow
$\propto \Sigma_{\rm gas}^{9/4}$ increases more rapidly with gas surface density
than the Eddington flux $\propto \Sigma_{\rm tot}$ so the system would likely
adjust to have $\Sigma_{\rm gas} \sim \Sigma_{\rm crit}$.  However, the required
inflow rates are extremely large, $\dot M \sim 2000 M_8^{3/4} M_\odot
\, {\rm yr^{-1}}$ at $\Sigma_{\rm gas} \sim 10^{11} \, {M_{\odot}}{\rm
  kpc^{-2}}$, and must be present even at very small radii ($\sim$ pc
for star clusters and $\lesssim 10-100$ pc for massive galaxies).
These required inflow luminosities are significantly larger than we
have found in a preliminary analysis of our numerical simulations of
the mergers of very massive gas-rich galaxies.  As a result, it
appears unlikely that the high inflow luminosities required to reach
the dust Eddington luminosity can in fact be realized in all of the
systems considered in Figure \ref{fig:profiles}.

One concern about invoking the dust Eddington limit is that the GCs in
Figures \ref{fig:mass.radius} \& \ref{fig:profiles} have low
metallicities, typically $\sim 0.1$ solar, but sometimes even lower.
As a result the dust opacity will be $\sim 10$ times smaller than that
used in equation (\ref{edd1}) and the massive stars present during GC
formation would not reach $\sim F_{\rm Edd}$.  We note, however, that
the winds from massive stars have a momentum flux comparable to that
of the photons (e.g., \citealt{starburst99}) and could have
an impact similar to that attributed to photons here.

%Another concern is $N$-body relaxation and gravothermal collapse -- 
%the profiles considered here may not reflect those at formation. 
%However, the nature of gravothermal collapse is such that it should 
%run away, not halt at some characteristic maximum. 
%Considering the relaxation time $t_{\rm R}\sim 0.1\,N\,\ln^{-1}N\,t_{\rm dyn}$, 
%for a virialized system at surface density near $\Sigma_{\rm max}$, 
%we note $t_{\rm R}\sim 10^{10.5}\,{\rm yr}\,
%[\Sigma_{\rm tot}/\Sigma_{\rm max}]^{1/2}
%[R/10\,{\rm pc}]^{5/2}$. Note the steep dependence on $R$; for 
%GCs, it is well-known that $t_{\rm R}<10^{9}\,$yr, implying they 
%may be affected; but relaxation cannot explain the 
%$\gtrsim10\,$pc scales on which the centers of galaxies, 
%spheroidal nuclei, and UCDs reach $\Sigma_{\rm max}$. 
%Caution is needed, however, at sub-pc scales, where this time is 
%very short ($\sim10^{8}\,$yr). There may be features in e.g.\ the profiles of core collapse 
%clusters reflecting this collapse, that could greatly exceed $\Sigma_{\rm max}$. 
%It is unlikely, even in such cases, 
%however, that larger scales (in particular those compared here) 
%are strongly modified. Simulations of core collapse \citep{cohn:core.collapse,chernoff:core.collapse}
%tend toward sharp nuclear ($\ll$pc) mass density concentrations 
%and stellar mass segregation, but the 
%mass fraction shifted is small, and the surface mass density on scales $\sim$pc 
%(the range of interest here) is largely unaffected 
%(changing by factors $\lesssim2-3$, within the scatter seen here). 

Another concern about our conclusions drawn here is that the profiles
observed now may not reflect those at formation, because of $N$-body
relaxation.  The relaxation time for a virialized system near
$\Sigma_{\rm max}$ is $t_{\rm R}\sim 10^{10.5}\,{\rm yr}\,
[\Sigma_{\rm tot}/\Sigma_{\rm max}]^{1/2} [R/10\,{\rm
 pc}]^{5/2}$. Note the steep dependence on $R$; for GCs, it is
well-known that $t_{\rm R}<10^{9}\,$yr, implying they may be
significantly affected by relaxation; but relaxation cannot explain
the $\gtrsim10\,$pc scales on which the centers of galaxies,
spheroidal nuclei, and UCDs reach $\Sigma_{\rm max}$.  Caution is
needed, however, at sub-pc scales, where the relaxation time is very
short ($\sim10^{8}\,$yr).  Even in such cases, however, it is unlikely
that the somewhat larger scales considered here are strongly
modified. Simulations of core collapse
\citep{cohn:core.collapse,chernoff:core.collapse} tend toward sharp
nuclear ($\ll$pc) mass concentrations and stellar mass segregation,
but the mass fraction shifted is small, and the surface mass density
on scales $\sim$pc (the range of interest here) is largely unaffected
(changing by a factor of $\lesssim2-3$, within the scatter seen here).

The maximum stellar surface density found here could readily be exceeded if
gas ``trickled in'' slowly from large radii, and was allowed to form
stars for a long time and at a low rate.  In principle, there should
be no limit to how large a stellar surface density might be attained
in this case because all forms of feedback would be ineffective at low
star formation rates; in particular, the system would remain well
below the Eddington limit.  The interpretation given here requires
that the high-density portions of the galaxy or cluster form in no
more than a few, rapid events; this appears consistent with other
observational indicators (e.g.\ the stellar populations).

\vspace{-0.33in}
\acknowledgments 
We thank Tod Lauer and Kevin Bundy for helpful discussions and the
Aspen Center for Physics, where a portion of this work was conceived.
Support for PFH was provided by the Miller Institute for Basic Research 
in Science, University of California Berkeley.
EQ is supported in part by NASA grant NNG06GI68G and 
the David and Lucile Packard Foundation.  TAT is supported
in part by an Alfred P.~Sloan Fellowship.
\\

\bibliography{/Users/phopkins/Documents/lars_galaxies/papers/ms}
%\bibliographystyle{mn2e}
%\bibliography{max_densities}

\end{document}